\documentclass[%
 aip,
 amsmath,amssymb,
 reprint,%
]{revtex4-1}

\usepackage{graphicx}% Include figure files
\usepackage{dcolumn}% Align table columns on decimal point
\usepackage{bm}% bold math
\usepackage[utf8]{inputenc}
\usepackage[T1]{fontenc}
\usepackage{mathptmx}
\usepackage{etoolbox}

\makeatletter
\def\@email#1#2{%
 \endgroup
 \patchcmd{\titleblock@produce}
  {\frontmatter@RRAPformat}
  {\frontmatter@RRAPformat{\produce@RRAP{*#1\href{mailto:#2}{#2}}}\frontmatter@RRAPformat}
  {}{}
}%
\makeatother
\begin{document}

\preprint{AIP/123-QED}

\title{Position fixing with cold atom gravity gradiometers}

\author{Alexander M. Phillips}
 \email{A.M.Phillips@liverpool.ac.uk}
 \affiliation{Department of Electrical Engineering and Electronics, University of Liverpool,  Brownlow Hill, Liverpool, L69 3GJ, UK.}
\author{Michael J. Wright}
 \email{M.J.Wright@liverpool.ac.uk}
 \affiliation{Department of Electrical Engineering and Electronics, University of Liverpool,  Brownlow Hill, Liverpool, L69 3GJ, UK.}
\author{Isabelle Riou}
 \email{Isabelle.Riou@Teledyne.com}
 \affiliation{Teledyne e2v, Chelmsford, CM1 2QU, UK.}
\author{Stephen Maddox}
 \email{maddoxs@gmail.com}
 \affiliation{Teledyne e2v, Chelmsford, CM1 2QU, UK.}
\author{Simon Maskell}
 \email{smaskell@liverpool.ac.uk}
 \affiliation{Department of Electrical Engineering and Electronics, University of Liverpool,  Brownlow Hill, Liverpool, L69 3GJ, UK.}
\author{Jason F. Ralph}
 \email{jfralph@liverpool.ac.uk}
 \affiliation{Department of Electrical Engineering and Electronics, University of Liverpool,  Brownlow Hill, Liverpool, L69 3GJ, UK.}

\date{\today}

\begin{abstract}

This paper proposes a position fixing method for autonomous navigation using partial gravity gradient solutions from cold atom interferometers. Cold atom quantum sensors can provide ultra-precise measurements of inertial quantities, such as acceleration and rotation rates. However, we investigate the use of pairs of cold atom interferometers to measure the local gravity gradient and to provide position information by referencing these measurements against a suitable database. Simulating the motion of a vehicle, we use partial gravity gradient measurements to reduce the positional drift associated with inertial navigation systems. Using standard open source global gravity databases, we show stable navigation solutions for trajectories of over 1000km. 
\end{abstract}

\maketitle

\section{Introduction}
\label{sec:sec1}

Cold atom sensors offer unprecedented precision and accuracy in the measurement of inertial quantities~\cite{Gustavson1997,Gustavson2000,Canuel2006,Geiger2011,Dickerson2013,Wu2017,Yankelev2019,Garrido2019}, including gravitational acceleration~\cite{Niebauer1995,Merlet2010,Canuel2018,Bidel2020} and gravity gradients~\cite{Snadden1998,McGuirk2002,Sorrentino2012,Biedermann2015,Veryaskin2018,Janvier2022,Bongs2019,Stray2022,Janvier2022}. The accuracy of cold atom devices derives from a quantum mechanical property of atoms, the phase sensitivity of matter wave interferometry~\cite{Scully1993,Dowling1998}, which can be several orders of magnitude (or more) better than optical interferometers, and the fact that the masses of all isotopically pure atoms are all identical, meaning that systematic biases in gravity measurement due to minute differences in mass are effectively zero. 

Whilst the ability to measure inertial quantities very accurately may ultimately improve inertial navigation systems (INSs), the technology is still relatively immature~\cite{Garrido2019,Jekeli2005,Fang2012,Cheiney2018} when compared to classical inertial sensors and navigation systems~\cite{Titterton1997,Groves2013,Nebylov2016}. However, good progress is being made to take cold atom systems out into the real world and to turn them into practical sensors~\cite{Geiger2011,Wu2017,Garrido2019,Bongs2019,Stray2022}. 

Cold atom quantum inertial navigation, like classical inertial navigation, is a form of dead reckoning, in that it measures accelerations and angle rates to obtain estimates of the desired quantities: position, velocity and attitude or orientation. Even highly accurate quantum sensors will still suffer from the tendency of all dead reckoning systems to drift over time as errors accumulate~\cite{Jekeli2005}. Navigational drift would be present, even if the sensors were perfect. 

The way to correct the tendency of an inertial navigation system to drift is to augment it with a position fixing system, which measures position directly rather than by integration (twice) of acceleration measurements. The most common form of position fixing is a Global Navigation Satellite System (GNSS), such as the US Global Positioning System (GPS), which provide position information, together with Doppler velocity and timing signals~\cite{Groves2013}. However, these signals suffer from well-known vulnerabilities to intentional jamming and spoofing~\cite{Schmidt2017}. To provide alternatives to GNSS position fixing, a range of other methods have been developed. These alternatives include terrain referenced navigation (using laser or radar sensors)~\cite{Groves2006}, camera-based visual map-matching~\cite{Moreira2019,Bijjahalli2020}, and terrestrial radio signals; which could be navigation specific (e.g. LORAN/e-LORAN~\cite{Son2017}) or standard broadcast signals (often termed `signals of opportunity')~\cite{Kapoor2017}. Such technologies offer an alternative to GNSS by referencing the position of the vehicle against a database of collateral features present in the environment. In the absence of suitable external features, position fixing will not work. For example, for navigation over the sea, there are no surface features (terrain or visual) to correlate against, and for underwater vessels it may be disadvantageous to use active sonar to measure features on the seabed because it gives away your own position. Terrestrial radio signals are sensitive to jamming and spoofing, although this would require broad spectrum jamming/spoofing, which is more complicated than it would be for low power and relatively narrow bandwidth GNSS signals. 

Cold atom gravity sensors offer an alternative to existing feature based methods, using measurements of the variations in the strength of the gravitational field in different locations and correlating the measured gravitational features against a suitable database~\cite{Gleason1995,Jekeli2006,Welker2013,Wang2016,Wang2017,Phillips2020}. Gravitational features are present in all regions of the globe, over both land and sea, and they are impossible to jam or to spoof. This, together with the ubiquitous nature of the gravity signal, means that position fixing is autonomous and it is entirely passive. Gravitational databases are freely available for all areas of the globe; for example, a global database for gravitational variations associated with measurements of the Earth's geoid (EGM2008)~\cite{Pavlis2012} is available online and it has a resolution of 1 nautical mile -- the geoid maps the local variations of gravity to allow local mean sea level to be defined relative to the standard Earth reference ellipsoid, WGS84~\cite{Merrigan2002}. Higher resolution databases are available for most of the land masses. For example, SRTM2gravity~\cite{Hirt2019}, which is modelled from local topographic features, currently has a resolution of 90 metres. 

The main difficulty with position fixing using gravitational map-matching is that the variations in gravity are very small and often difficult to measure. Measuring variations in the acceleration due to gravity is possible with sensitive classical accelerometers, using sensitive optical measurements of falling bodies (`falling cube' gravimeters)~\cite{Niebauer1995,Merlet2010}. Gravitational map-matching has already been demonstrated using such systems~\cite{Wang2016,Wang2017}. Cold atom quantum interferometer systems offer significant scope for practical gravity-based map-matching systems, and -- unlike conventional classical gravimeters -- they do not require frequent calibration to retain their accuracy over long periods of time. 

This paper proposes a method for position fixing using partial measurements from a gravity gradient sensor formed from two cold atom interferometers with a common laser reference signal, which allows the cancellation of vibrational noise between the two measurements. The standard method for extracting gravity gradient measurements using this type of sensor is to take a sequence of paired measurements from each interferometer. The pairs of measurements lie on an ellipse -- because the phase of the reference laser is unknown and varies over time. The gradient is found by fitting an ellipse to a set of measurements~\cite{Foster2002,Stockton2007,Xu2018}. This presents a potential problem for the use of this method when the gradiometer is moving relative to the reference field. If the gravity gradient varies between each pair of interferometer measurements, each pair lies on a different ellipse. As a result, variations in the gravitational gradient as the sensor moves introduce errors into the measured gradients. To avoid these errors, we propose a position fixing method that references individual pairs of interferometer measurements against a gravitational database. The method uses a set of possible locations and the gravitational database to produce a set of candidate ellipses, and uses a particle filter~\cite{Gustafsson2002,Aru2002,Cappe2007} to select locations that minimise the difference between the candidate ellipses and the actual measurement pairs. Over time, as the vehicle moves -- with each particle in the particle filter representing a location and an ellipse -- the distribution of particles will converge on the correct position. We describe this as position fixing with {\em partial} gravity gradient measurements, because we do not construct an ellipse over a sequence of measurement pairs in the conventional way, the ellipses that we construct correspond to candidate locations and the pairs of interferometer measurements are used as independent updates to improve the estimated position. We will however compare the results against the standard ellipse fitting method, which shows a small bias in the gravity gradient values obtained underestimating the true gravity gradients experienced.

The paper starts in section~\ref{sec:sec2} by introducing the measurement model used for the gravity gradiometer. Section~\ref{sec:sec3} outlines the inertial navigation solution and the vehicle trajectory model and section~\ref{sec:sec4} describes the particle filter for the position fixing updates. Example results are presented in section~\ref{sec:sec5} and a discussion of the results and conclusions are given in section~\ref{sec:sec6}.

\section{Cold atom gravity gradiometers}
\label{sec:sec2} 

Cold atom gravity gradiometers consist of two atom interferometers separated by a known distance and sharing a common Raman laser reference signal~\cite{Snadden1998,McGuirk2002,Sorrentino2012,Biedermann2015,Veryaskin2018,Janvier2022,Bongs2019,Stray2022,Janvier2022}. Clouds of cold atoms are prepared and held in a magneto-optical trap inside a vacuum chamber. During each measurement, the clouds in each chamber are allowed to fall under the action of gravity. During the fall, the laser provides a sequence of pulses for each cloud: a $\pi/2$ pulse is applied to place the atoms in a superposition of momentum states (the interferometer `beam-splitter'), the second pulse (a $\pi$ pulse) reverses the momentum states in the superposition (`reflection'), and another $\pi/2$ pulse recombines the superpositions, which have acquired a phase difference due to the different trajectory taken by each of the superposed states. The phase difference causes interference fringes which can be measured and related to the gravitational acceleration experienced by the atoms. For this paper, we are interested in the gradient of the vertical component of the gravitational acceleration so we place the interferometers one above the other, and on a stabilised platform, to measure the vertical component of the gravitational acceleration at two points separated by a vertical distance $\Delta z$. Other configurations have been proposed for measuring other components of the gravity gradient and differential measurements~\cite{Biedermann2015} but these are not considered here.

The sensor measurements are simulated using a standard model to provide two signals, one for each of the interferometers~\cite{McGuirk2002,Sorrentino2012},
\begin{eqnarray}
S_0(n) &=&  \eta (\bar{N}+\delta N_n) \sin(\phi_0 + \phi_n) + s_0 \label{signals1}\\
S_1(n) &=& \eta (\bar{N}+\delta N_n) \sin(\phi_1 + \phi_n + \delta\phi_n) +s_1 \label{signals2}
\end{eqnarray}
where $\eta$ is the measurement efficiency (or the contrast of the interference fringes), $\bar{N}$ is the average number of atoms in each cloud, $\delta N_n$ is the atom number shot noise on each measurement (at time step/measurement $n$, where the standard deviation of $\delta N_n$ is $\sigma_N = \sqrt{\bar{N}}$), $\phi_n$ is the Raman laser phase, which is unknown but common to each interferometer in a gradiometer, $\delta\phi_n$ is the measurement phase noise which is assumed to be Gaussian with a standard deviation $\sigma_{\phi}$ (both interferometer would have such phase noise, but the phase noise in the upper `0' interferometer is subsumed within the definition of $\phi_n$), and $s_0$ and $s_1$ are both constants representing signal biases. $\phi_0$ and $\phi_1$ are the phases of interest in each of the interferometers, $\phi_0 = k_{eff} g(z_0) T^2$ and $\phi_1 = k_{eff} g(z_1) T^2$, where $k_{eff} $ is the effective wave number of the $\pi/2$ interferometer pulses, and $T$ is the time between the $\pi/2$ and the $\pi$ pulses. For convenience, we will normalise signals by dividing through by $\eta\bar{N}$: $\tilde{S}_{0,1}(n) = S_{0,1}(n)/ \eta\bar{N}$.

Taking the difference of these two phases provides the value of the gradient of the vertical component of the gravitational acceleration in the vertical direction $dg_z/dz$. 
\begin{eqnarray}\label{phases}
\Delta \phi &=&  \phi_0 - \phi_1 \\
&=& k_{eff} (g(z_0) - g(z_1)) T^2 \\
&\simeq& k_{eff} \Delta z (dg_z/dz) T^2
\end{eqnarray}
However, the unknown phase in the Raman laser means that each pair of measurements lies on an ellipse that can be formed by taking repeated measurements and fitting an ellipse to the resultant signals, as described in reference~\cite{Foster2002}. Often 20 or more measurements are required for to obtain a good fit to the ellipse. More sophisticated sensor models that include other effects -- including rotations and higher order gravity terms -- are available~\cite{Peters2001,Sorrentino2014,Bertoldi2019} but the standard approach is adopted here for simplicity and to focus attention on the position fixing method. 

The standard ellipse fitting approach described in reference~\cite{Foster2002} has been extended and enhanced~\cite{Stockton2007,Xu2018}. However, again for simplicity, we use the standard method where the phase difference is found from fitting an ellipse of the form~\cite{Foster2002},
$$
A x^2 + B xy + C y^2 + D x + E y + F = 0
$$
so that $\Delta\phi = \cos^{-1}(-B/2\sqrt{AC})$. The two constants, $s_0$ and $s_1$, found in~(\ref{signals1}) affect the origin of the fitted ellipse but not the phase difference that we use to estimate the gravity gradient, so we set these to zero for convenience. 

For the sensors modelled in this paper, the relevant cold atom sensor parameters values are: interferometer measurement frequency $f_{meas} = 1.0$ Hz, atomic mass (Caesium 133) $m = 2.20693925\times 10^{-25}$ kg, time between pulses $T = 0.16$ s, atom recoil velocity $v_{rec} = 7.0 \times 10^{-3}$ m/s, $k_{eff} = m v_{rec}/\hbar$, vertical separation of interferometers $z_0 - z_1 = \Delta z = 0.5$ m, mean number of atoms per measurement $\bar{N} = 10^{6}$, measurement efficiency $\eta = 0.5$. The specific choices for the interferometer parameters are based on system parameters from the current generation of cold atom technologies, and they are not critical for the operation of the position fixing algorithm discussed in this paper. 

\section{Inertial navigation and trajectory modelling}
\label{sec:sec3} 

To demonstrate the proposed method, we use an aviation example, where we simulate a cold atom sensor placed in a large aircraft. The aircraft is assumed to have a standard classical inertial navigation system that provides high frequency motional data and a navigation solution that will drift if unaided by the cold atom sensor and the position fixing provided by the method presented here. The aircraft flies in a straight line between two points at a constant altitude. The two points are selected to be sufficiently far apart (more than 1000 km) so that the navigation solution has to account for a number of important factors, including the variations in the local gravity field. In addition to gravitational variations, simulating a trajectory over a large distance and a long duration flight we must also include the curvature of the Earth and -- travelling North to South -- the approximately ellipsoidal shape of the Earth, the Coriolis effect due to the Earth's rotation, and the requirement to change the local navigation frame to maintain a local level (the `transport rate'~\cite{Groves2013}). 

\begin{figure}[t] 
   \centering
   \includegraphics[width=1.0\hsize]{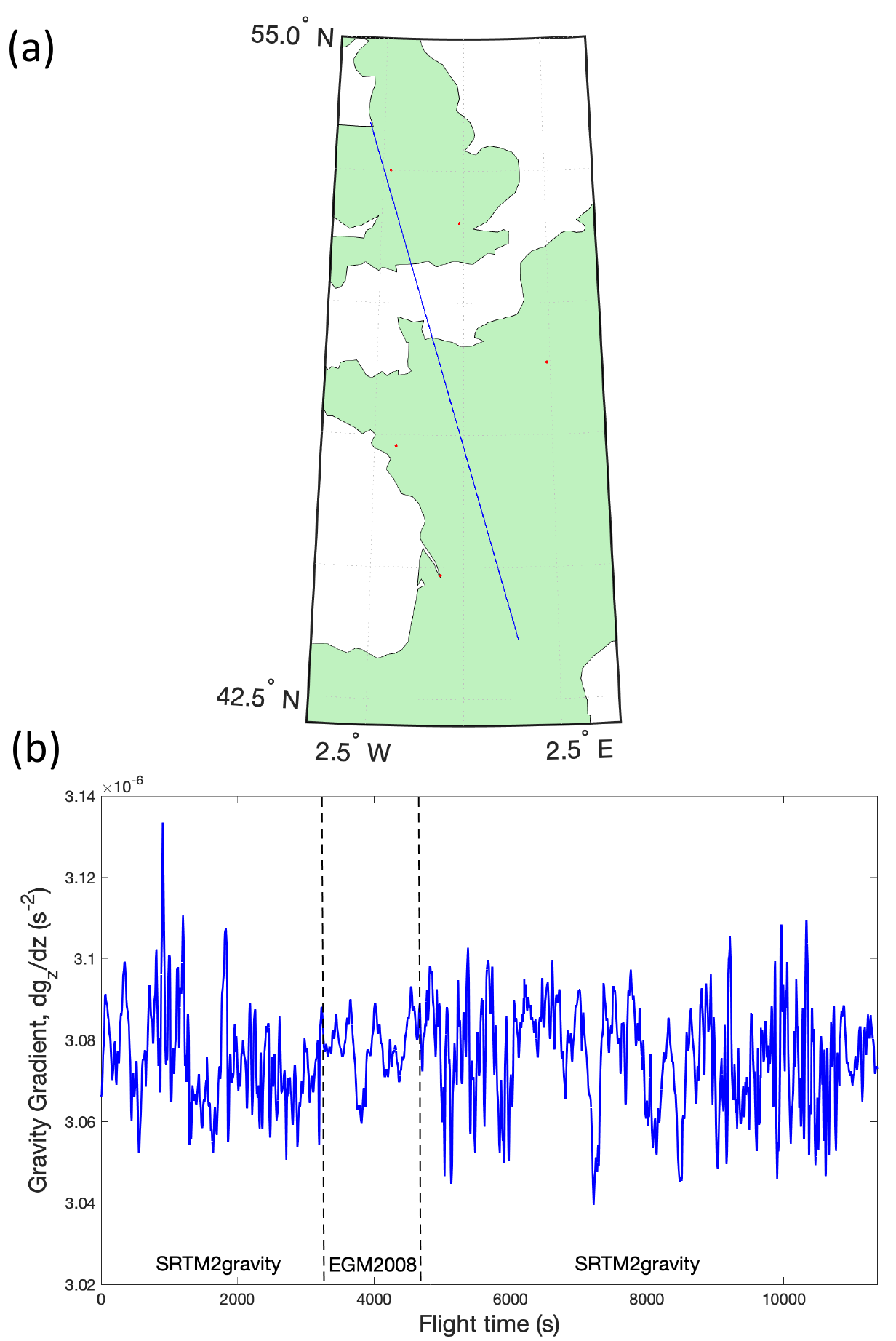} 
   \caption{(a) Simulated route -- Liverpool to Toulouse -- and (b) simulated gravity gradients, showing regions where the SRTM2gravity and the EGM2008 databases are used.}
   \label{trajectory}
\end{figure}

For this paper, we model a trajectory that starts in Liverpool (Latitude = $53.407579$ degrees, Longitude = $-2.967853$ degrees) and ends in Toulouse (Latitude = $43.604652$ degrees, Longitude = $1.444209$ degrees), which corresponds to a distance of approximately 1137 km. The aircraft flies at a constant altitude of 3000m and at a constant speed of 100 m/s, meaning that the total journey takes 11370 seconds or 3 hours and 9.5 minutes. We have chosen this route for the reasons outlined above and because the route passes over a body of water -- the English Channel or la Manche. This forces us to use two gravity databases, a high resolution SRTM2gravity database which is only defined for sections of the route over land (with a 90m resolution~\cite{Hirt2019}) and the EGM2008 global database (with a resolution of 1 nautical mile~\cite{Pavlis2012}) over the sea area: the simulation has been set up to use SRTM2gravity where it is available and default to EGM2008 where it is not available. The gravity gradient values are calculated using the numerical integration method described in reference~\cite{Hofmann2006}. Using this route allows us to investigate the effect of different database resolutions on the accuracy of the resultant navigation solution. The route and the simulated gravity gradient profile generated by the two databases are shown in Figure~\ref{trajectory}. The background gravity gradient is $dg_z/dz \simeq 3.07\times 10^{-6} {\rm s}^{-2}$, whilst the local variations that are used in position fixing are around $2-3\times 10^{-8} {\rm s}^{-2}$. The sections of the route corresponding to the higher resolution SRTM2gravity database show significantly larger variability with more smaller scale features. 

The trajectory is defined in terms of a series of equally spaced waypoints linking the two locations (spaced at 1 second intervals along the path) and, for simplicity, we assume that the motion is smooth and free of large vibrations and sudden jerks -- although we do include small vibrations, which can cause atom clouds to drift outside the Raman beam and cause a measurement to fail. We also assume that the cold atom sensor is on a stabilised platform to remove the effect of small rotations of the platform~\cite{Bidel2020}. We interpolate positions between the waypoints and convert these locations to a local navigation frame~\cite{Groves2013}, which is defined in terms of local North-East-Down axes in this work. The definition of a local Cartesian reference frame (North-East-Down or NED, often referred to as the local Earth-oriented frame) is standard practice in aerospace engineering and simplifies the generation of motional states: velocity, acceleration, and rotational states/angular velocities. We assume that the aircraft orientation is fixed relative to the platform's velocity vector so that the angles of attack and sideslip (the angles between the platform body axes and the velocity vector) are constant, and we can, without loss of generality, set these to zero. 

This approach, together with a WGS'84 ellipsoid and gravity mode, allows us to generate all of the relevant dynamic quantities (positions, velocities, accelerations, attitude angles, angle rates) at whatever frequency is required for the navigation system model. We then use the standard equations for inertial navigation in the local frame, described in detail in reference~\cite{Groves2013}, to calculate an inertial navigation solution for the route that we have defined. We assume a perfectly aligned navigation solution initially and use standard values for the standard deviation of errors present in aviation grade inertial navigation systems~\cite{Groves2013} -- including accelerometer/gyroscope fixed biases, accelerometer/gyroscope non-orthogonalities, accelerometer/gyroscope scaling errors, and accelerometer/gyroscope measurement errors, see Table~\ref{aviationgradeerrors} for the error values used in this paper. We generate simulated measurements and an open loop INS solution that drifts over the course of the flight. This open loop INS solution provides the baseline case for the performance of an unaided INS without access to a position fixing system.
\begin{table}[h!]
\begin{center}
\begin{tabular}{| l | l | } 
 \hline
Sensor Error & Error Value (1 std dev.)  \\  
 \hline\hline
 Accelerometer Static Bias & 30 micro-g \\
 \hline
 Accelerometer Non-Orthogonality &  10 micro-radians\\
 \hline
 Accelerometer Scaling Error & 10 ppm\\
 \hline
 Accelerometer Measurement Noise & 15 micro-g/$\sqrt{\rm Hz}$ \\ 
 \hline\hline
 Gyroscope Static Bias & 0.05 micro-radians \\
 \hline
 Gyroscope Non-Orthogonality &  10 micro-radians\\
 \hline
 Gyroscope Scaling Error & 10 ppm\\
 \hline
 Gyroscope Measurement Noise & 2.0 micro-rad/sec.$\sqrt{\rm Hz}$ \\ 
 \hline
\end{tabular}
\caption{Error values used for the INS sensor simulations.}
\label{aviationgradeerrors}
\end{center}
\end{table}
Most high grade inertial systems operate at very high frequency, normally several hundred Hz, but for the simulations here we simulate the navigation solution running at a frequency of 100Hz, which is sufficiently high to accurately navigate along the relatively benign trajectory that we have defined, but low enough to allow us to simulate a number of runs for each of the cases discussed below within a reasonable timescale. The navigation solution is integrated, as described in reference~\cite{Groves2013}, with time steps $\delta t = 0.01$ seconds.

We define the state vector for the navigation solution at time $t$, which includes position, velocity, acceleration, attitude angles (Euler angles) and angle rates by,
$$
\underline{X}_t = \left(
\begin{array}{c}
\Phi \\
\Lambda \\
h \\
u \\
v \\
w \\
a_x \\
a_y \\
a_z \\
\psi \\
\theta \\
\phi \\
P \\
Q \\
R 
\end{array}
\right)
= \left(
\begin{array}{l}
{\rm latitude, degrees} \\
{\rm longitude, degrees}  \\
{\rm altitude, metres}  \\
x\:{\rm velocity, body\:axes, m/s}  \\
y\:{\rm velocity, body\:axes, m/s} \\
z\:{\rm velocity, body\:axes, m/s} \\
x\:{\rm acceleration, body\:axes, m/s^2} \\
y\:{\rm acceleration, body\:axes, m/s^2} \\
z\:{\rm acceleration, body\:axes, m/s^2} \\
{\rm platform\:heading, degrees} \\
{\rm platform\:pitch, degrees} \\
{\rm platform\:roll, degrees} \\
{\rm angle\:rate, body} \:x\: {\rm axis, deg/sec} \\
{\rm angle\:rate, body} \:y\: {\rm axis, deg/sec} \\
{\rm angle\:rate, body} \:z\: {\rm axis, deg/sec} 
\end{array}
\right)
$$
The position is defined in terms of the latitude, longitude and altitude of the platform relative to the WGS'84 ellipsoid and the attitude/Euler angles are defined as rotations relative to the local NED reference frame. The time derivatives are defined relative to the current body axes. Although the combination of different reference systems may seem complicated, this combination is a result of the need to navigate along routes where the Earth's curvature and rotation generate significant issues if one tries to approximate this with a single local Cartesian reference system. It is possible to use a single Cartesian reference system if one steps off the Earth and uses a single global Earth-Centred Inertial (ECI) reference frame or a rotating Earth-Centred Earth-Fixed (ECEF) frame~\cite{Groves2013}. The ECI and ECEF frames do simplify some of the calculations for the navigation solution, but they are less intuitive and can result in computational issues with small differences between very large quantities. 

Of the three translational degrees of freedom, the vertical INS position is normally the most unstable. This is because accelerometers measure the specific force rather than acceleration, which contains components from the gravitational acceleration as well as any acceleration due to motion of the platform. As a result, the acceleration measurements need to be corrected for the effect of gravity by subtracting an estimate of the local gravity. If the local gravity is incorrect or the position estimate is incorrect, the wrong gravity compensation will be applied in the navigation processing, leading to an instability in the vertical position that is not present in the horizontal position. To avoid this complication, we assume that the aircraft is fitted with an altimeter, which provides a good estimate of current altitude throughout the flight. Such altimeters are very common on aircraft so this is not an unreasonable assumption.

\section{Integrated gravity gradiometer-inertial navigation}
\label{sec:sec4} 
 
As discussed, an inertial navigation system on its own will drift over time, so we aim to use a sequence of simulated gravity gradient measurements to provide position fixes that will limit the extent of this drift. The method that we propose constrains the drift in position but can also limit the drifts in velocity and attitude.

To combine the high frequency INS solution represented by this state vector with the lower frequency data from the cold atom sensor, we define a correction vector, $\Delta\underline{X}(t)$, which contains corrections to the INS navigation solution in the local NED frame. Because the corrections are applied over much shorter timescales than the full trajectory, and we will be adding noise to these vectors at a later stage which will offset any issues arising from a local Cartesian approximation. We also limit the position corrections to the horizontal plane, because there is an altimeter to maintain a stable altitude estimate. 
$$
\Delta\underline{X}^{(NED)}_t = \left(
\begin{array}{c}
\Delta x \\
\Delta y \\
\Delta u \\
\Delta v \\
\Delta w \\
\Delta\psi \\
\Delta\theta \\
\Delta\phi 
\end{array}
\right)
= \left(
\begin{array}{l}
{\rm North\: correction,  NED, metres} \\
{\rm East\: correction, NED, metres} \\
x\:{\rm velocity\:correction, body, m/s}  \\
y\:{\rm velocity\:correction, body, m/s} \\
z\:{\rm velocity\:correction, body, m/s} \\
{\rm platform\:heading\:correction, deg} \\
{\rm platform\:pitch\:correction, deg} \\
{\rm platform\:roll\:correction, deg} 
\end{array}
\right)
$$

To construct a suitable correction vector, we use a particle filter~\cite{Gustafsson2002,Aru2002,Cappe2007}. Particle filters are general state estimation filters that are based on the idea of {\it sequential importance sampling} and are a type of sequential Monte Carlo (SMC) sampler~\cite{Cappe2007}. They have been applied to problems in map-matching for navigation before and have become popular for terrain referenced navigation~\cite{Gustafsson2002}. A particle filter uses a set of possible candidate solutions (the `particles') and probability to preferentially select out those solutions which best match the measured values. Each particle is weighted and re-weighted after each measurement using the probability that the measurement obtained could have come from that the solution for the corresponding particle. Between measurements, the state of each particle evolves according the dynamical evolution of the system, or more generally some prediction process represented by a proposal distribution. As more measurements are added, the particle weights will tend to cluster around the solutions which most agree with the measurements received. When the particle weights are concentrated on a subset of particles, the particles are resampled to move more particles into regions with high weight and away from the regions where particles have little weight. There are a number of ways to do particle resampling, but all rely on the distribution generated by the particle weights. Even particles with very low weight will normally have some chance of being selected as candidates even if that chance it relatively very low. In this way, a particle filter can gradually select good solutions over a series of measurements, and allow for the fact that the dynamic evolution itself may contain uncertainties. Additional noise can be added as `process noise' in the prediction step to reflect this uncertainty. The result is that a particle filter is normally robust to uncertainties and can handle significant nonlinearities and other features that tend to adversely affect other state estimation filters. 

Here, we use a simple type of particle filter, one called a `bootstrap' filter~\cite{Gustafsson2002,Aru2002,Cappe2007}, to estimate the correction to the inertial navigation solution. Each particle in the bootstrap filter corresponds to one candidate correction vector, $\Delta\underline{X}^{(NED),(i)}_t$, where $i = 1\ldots N_p$ and $N_p$ is the number of particles. Each candidate correction vector has a weight $w^{(i)}_t$, which are normalised after each measurement so that $\sum_{i=1}^{N_p} w^{(i)}_t = 1$. We set $N_p = 500$ in this paper, and resample using importance resampling~\cite{Cappe2007} when the effective number of particles $N_{eff}$ is less than a given threshold that is half of the total particle number, $N_{eff} = 1/(\sum_{i} (w^{(i)}_t)^2) < N_p/2$. 

The particles are re-weighted after each interferometer measurement corresponding to a pair of signals $\underline{S}_{meas} = (\tilde{S}_0, \tilde{S}_1)^T$, from (\ref{signals1}). To do this, we take the current INS solution and convert it to the local NED frame (i.e. the frame centred on the current navigation solution): $\underline{X}(t) \rightarrow \underline{X}^{(NED)}(t)$. We then add the $i$'th particle correction vector to generate a candidate navigation solution, for which we find the corresponding gravity gradient $dg^{(i)}_z/dz$ using the gravity database and the method described in reference~\cite{Hofmann2006}. This gravity gradient value is then used to generate a candidate ellipse, $\underline{S}^{(i)}$, by taking equations (\ref{signals1}), (\ref{signals2}) and (\ref{phases}) and letting the unknown phase $\delta\phi_n$ vary from $0$ to $2\pi$. We then calculate the minimum distance between the candidate ellipse and the pair of measurements taken from the interferometers, $\min |\underline{S}_{meas}-\underline{S}^{(i)}|$, and re-weight the particle using a Gaussian function
\begin{equation}\label{reweight}
\tilde{w}_{t}^{(i)}=\exp\left(- \frac{(\min |\underline{S}_{meas}-\underline{S}^{(i)}|)^2}{2 \sigma_S^2} \right) w_{t-\Delta t}^{(i)}
\end{equation}
where the $\tilde{w}^{(i)}$'s are the un-normalised weights after re-weighting and $\Delta t = 1$s $\gg\delta t$ is the time step between cold atom measurements. The standard deviation in the denominator in the exponential is the expected error in the signal measurements, which is a function of the interferometer noise parameters, $\sigma_S \simeq \sqrt{\sigma_N^2/\bar{N}^2+\sigma_{\phi}^2} = \sqrt{1/\bar{N}+\sigma_{\phi}^2}$. For the examples shown in this paper, the standard deviation is in the range $\sigma_S = 0.001$ to $0.010$. After all of the candidate ellipses have been calculated and all of the weights updated, the weights are renormalised and resampling is applied (if required). 

After re-weighting and resampling, we calculate the average correction vector $\Delta\underline{X}^{(NED)}_t $ to the INS solution in the local NED frame, which is given by,
$$
\Delta\underline{X}^{(NED)}_t = \sum_{i=1}^{N_p} w^{(i)}_t \Delta\underline{X}^{(NED),(i)}_t
$$
We then add this correction to the navigation solution provided by the INS and transform it back to the navigation frame: 
$$\underline{X}^{(NED)}(t) + \Delta\underline{X}^{(NED)}_t \rightarrow \underline{X}(t)$$
The particles/correction vectors are then re-centred on the current navigation solution: 
$$\underline{X}^{(NED),(i)}(t) \rightarrow \underline{X}^{(NED),(i)}(t)- \Delta\underline{X}^{(NED)}_t $$
In practice, we find that filtering the corrections gradually over several measurement cycles gives a more stable navigation solution, so we use a simple fixed gain filter to smooth the correction process,
$$\underline{X}^{(NED)}(t) + \alpha \Delta\underline{X}^{(NED)}_t \rightarrow \underline{X}(t)$$ 
and 
$$\underline{X}^{(NED),(i)}(t) \rightarrow \underline{X}^{(NED),(i)}(t)- \alpha\Delta\underline{X}^{(NED)}_t $$
where $\alpha= 0.05$ in the examples shown here.

The prediction step for the correction vectors can be done using the full navigation solution described in reference~\cite{Groves2013}. However, performing the calculations for each of 500 possible candidate solutions would be computationally expensive and, over a short period of time $\Delta t = 1$ second, a simple kinematic model is sufficient to provide an adequate update between the interferometer measurements. We propagate each of the correction vectors using the updated velocity and angle rates from the new navigation solution. To allow for uncertainties in this prediction step and to improve the robustness of the approach, we also add a small amount of process noise to the correction vectors: $\sigma_{pos} \simeq 25.0$ metres, $\sigma_{vel} \simeq 0.05$ metres/sec, and $\sigma_{att} \simeq 0.005$ degrees. The performance of the position fixing is not critically dependent on the levels of process noise added at this stage, as long as the noise level is sufficiently high to force the filter to explore enough of the neighbouring locations to maintain a `lock' onto the correct area of the gravity database, i.e. the particle filter covers enough of the area of the map to cover the correct area. 

Through this re-weighting and prediction process, the ellipses and the particles corresponding to these ellipses will be preferentially selected for higher weights. Rather than collecting a series of 20 or more pairs of interferometer measurements, which -- for a moving platform -- may come from areas of ground that have markedly different gravity gradients, the method presented here uses individual pairs of measurements and the position fixing is done in a continuous gradual process. We have deliberately used a simple variant of the particle filter to emphasise the robustness of the method presented in this paper. There are a large number of variants of particle filters and SMC samplers. It is likely that specific implementations of these may provide improved performance over the results presented here. However, the emphasis of this paper is on the general approach rather than the specifics of the filters used. 

\section{Results}
\label{sec:sec5}

We start by examining the ability of the proposed method to correct the position of the navigation solution where there is no phase noise, only shot noise. This provides a baseline case, and is motivated in part by recent work on systems that should be shot noise limited~\cite{Janvier2022}. We then consider the effect of phase noise on the approach and the robustness to failed measurements, which could be due to vibrations or other inadvertent motion causing the cold atoms to move outside the Raman beam. Although we assume that the interferometer sensor is mounted on a stabilised platform, a stabilisation system will not be able to compensate for all of the possible dynamical effects that an aircraft may experience. 
\begin{figure}[t] 
   \centering
   \includegraphics[width=1.0\hsize]{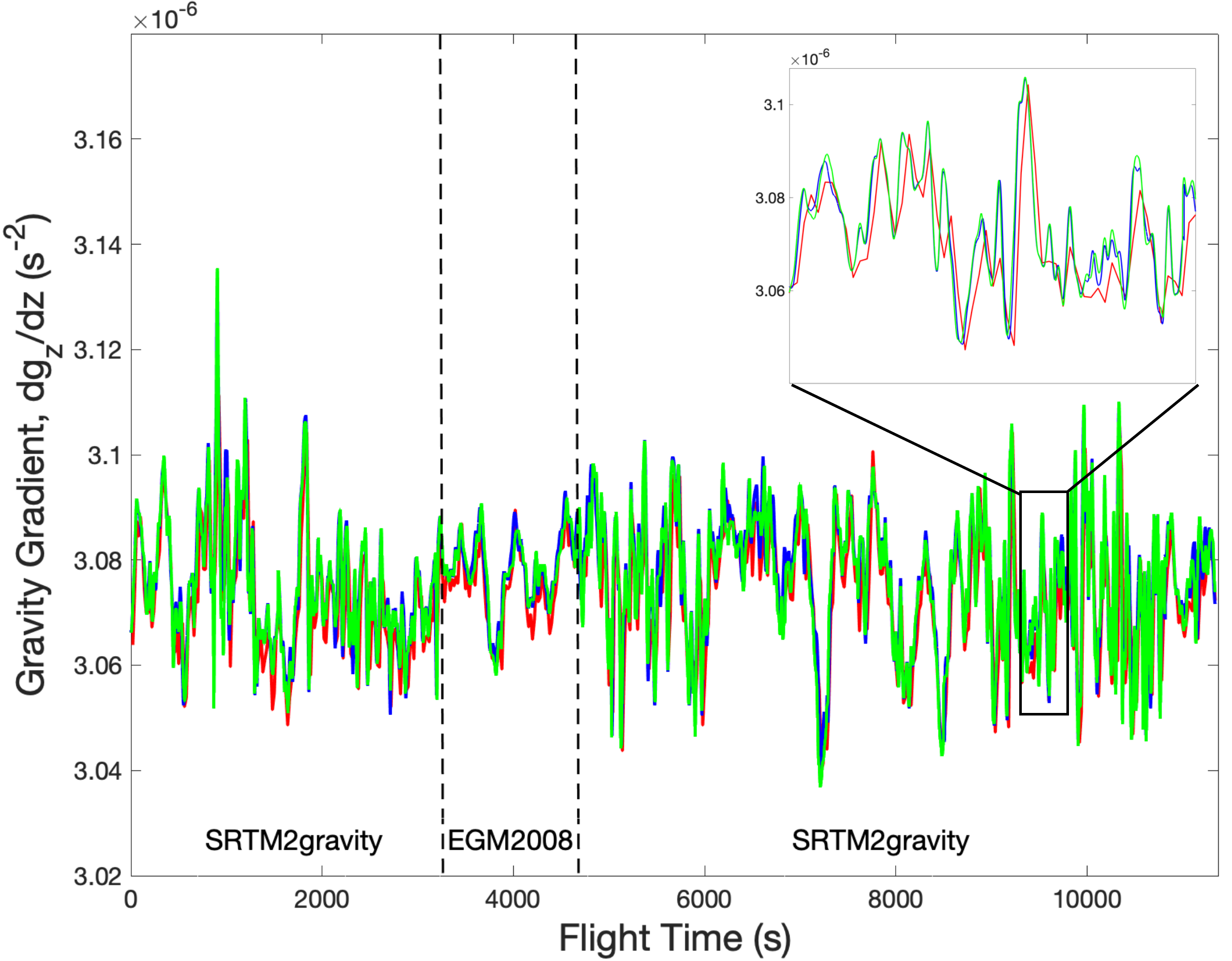} 
   \caption{Example of estimated gravity gradients using partial measurements (green) and using the conventional matched ellipse method (red), with simulated gravity gradient values (blue). A section of the profile is shown as an inset.}
   \label{gravity_gradient_example}
\end{figure}
\begin{figure}[h] 
   \centering
   \includegraphics[width=1.0\hsize]{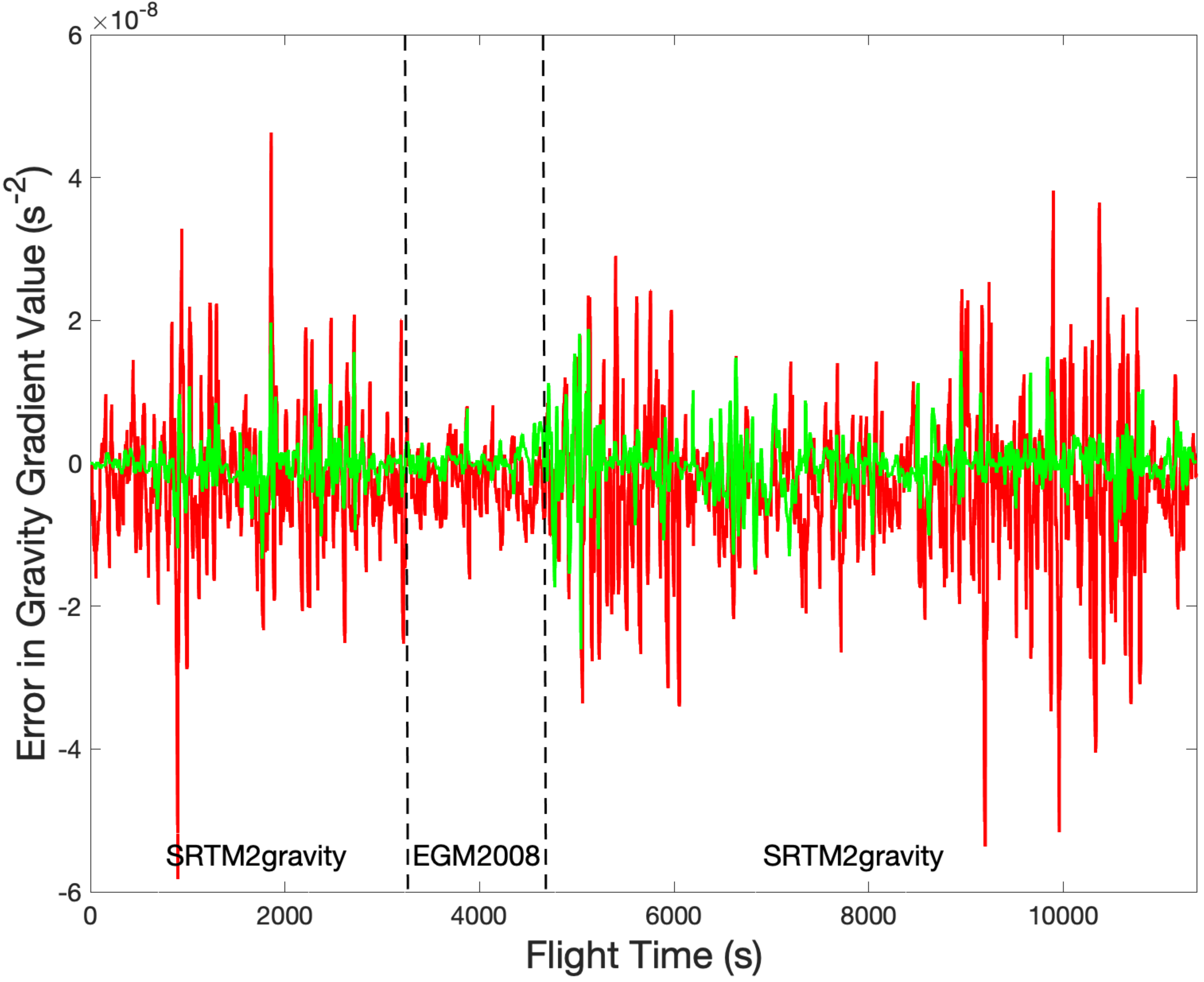} 
   \caption{Example of estimated gravity gradient errors using partial measurements (green) and using the conventional matched ellipse method (red).}
   \label{gravity_gradient_errors}
\end{figure}

\subsection{Zero Phase Noise}\label{sec5a}

Figure~\ref{gravity_gradient_example} shows an example of the estimated gravity gradients along the simulated trajectory for the gradient estimated from the database using the navigation solution generated by the position fixing method proposed in this paper. Strictly speaking, it is not a measurement of the gravity gradient in the true sense but it gives an indication of how well-matched the navigation solution is to the database. The figure also shows the gradient values found using the standard ellipse fitting method (in red). We also provide an insert with a small section of the gravity profile to show the differences between the two estimated solutions and the actual database values -- the estimates found from the partial gradient position fixing are found to be significantly better than the values found using the standard technique. We show the errors in the estimates for this example in Figure~\ref{gravity_gradient_errors}. The errors for the example shown using the standard ellipse fitting technique have a standard deviation of $9.4\times 10^{-9}$s$^{-2}$ and a mean value $-3.0\times 10^{-9}$s$^{-2}$, which shows a slight bias that underestimates the gravity gradient value. For the proposed method, the errors are smaller $3.6\times 10^{-9}$s$^{-2}$ and the mean value is an order of magnitude smaller $-1.7\times 10^{-10}$s$^{-2}$. 
\begin{figure}[t] 
   \centering
   \includegraphics[width=1.0\hsize]{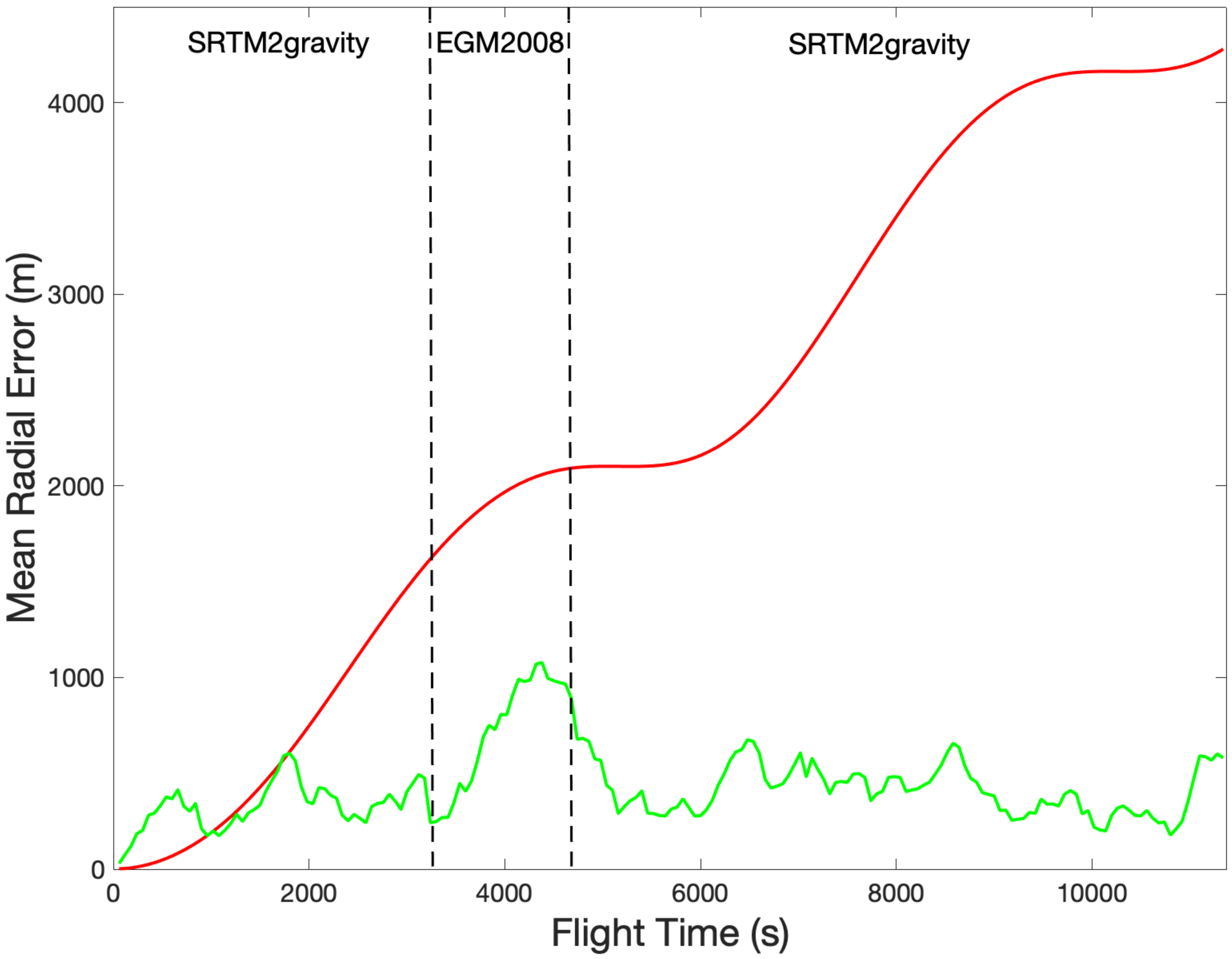} 
   \caption{Mean radial position error as a function of time for simulated flight, Liverpool-Toulouse, average taken over 50 flights: partial gravity gradient position fixing (green) and unaided INS without position fixing (red).}
   \label{radial_error_example}
\end{figure}

Figure~\ref{radial_error_example} shows the horizontal radial position error as a function of time averaged over 50 simulated flights. The unaided INS drifts over time, with the average positional drift being more than 4km after 3 hours of flight -- the oscillations in the unaided INS error are due to the Schuler oscillations, which are a well-known effect in long range navigation~\cite{Groves2013}. By contrast, the radial position errors for the INS with position corrections from the gravity matching method are limited to $\simeq 300-500$ metres for most of the route. The figure shows the sections of the trajectory that use the SRTM2gravity (over the land) and EGM2008 (over water). There is a marked degradation of performance in terms of the positional accuracy for the section of flight over the regions only covered by the EGM2008 database. This is due to the fact that the database features in this region do not have the same resolution as the SRTM2gravity database (1 nautical mile rather than 90m), and the gravity gradient has noticeably fewer small scale features that would help to localise the aircraft. Once the aircraft crosses the sea area, the positional accuracy improves quickly as the higher resolution features of the SRTM2gravity database are available again. There is some deterioration in the positional accuracy when the aircraft flies over central France, however, this also corresponds to a region where the gravity features are smaller relative to the areas where the localisation is better. In this case, over central France, the SRTM2gravity has fewer distinct features to align against. 

\subsection{Sensitivity to Phase Noise}\label{sec5b}

\begin{figure}[t] 
   \centering
   \includegraphics[width=0.9\hsize]{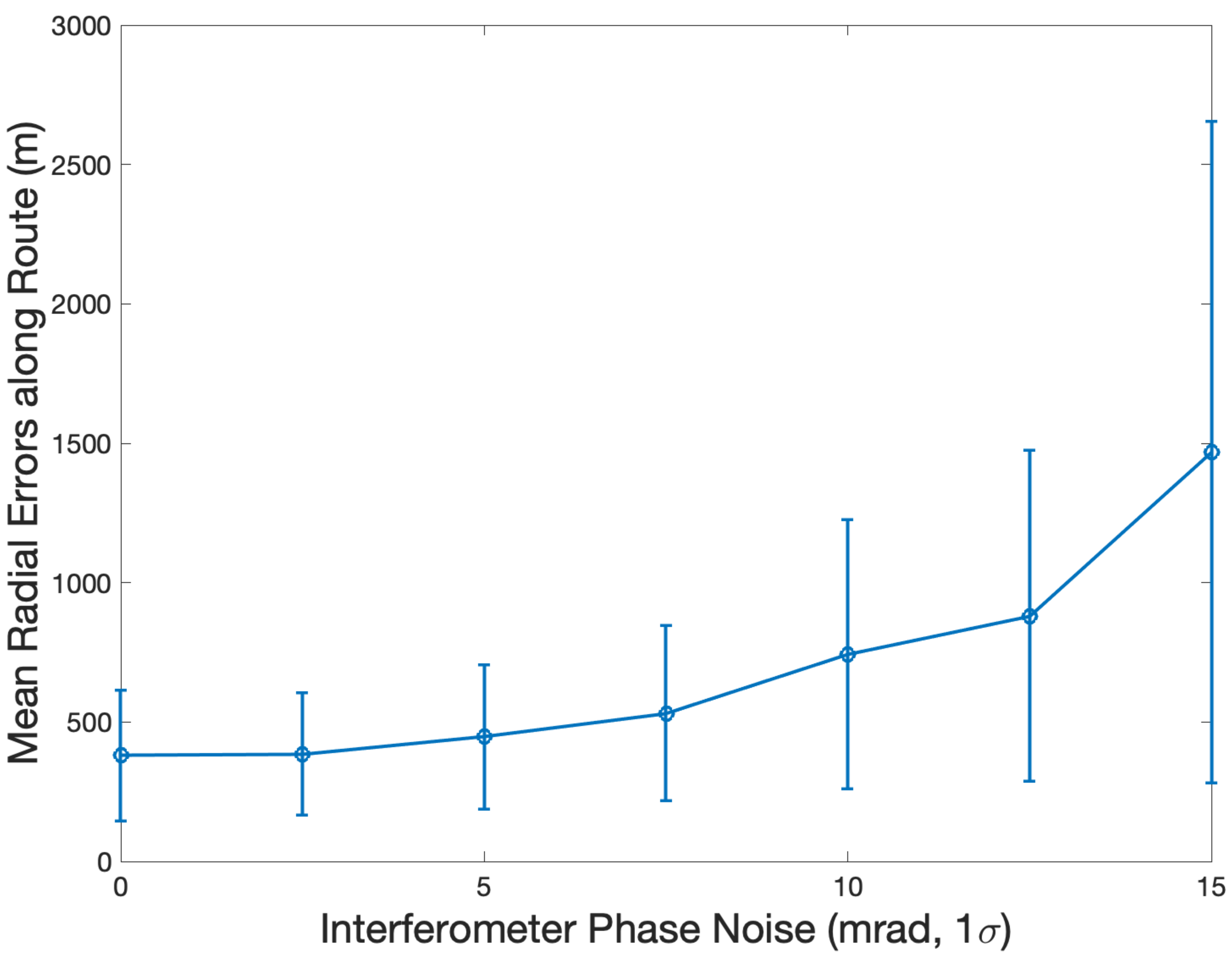} 
   \caption{Mean radial position error over 50 simulated flights and averaged along route as a function of the level of phase noise in the interferometers, with error bars ($1\sigma$).}
   \label{radial_error_vs_phasenoise}
\end{figure}

The accuracy of the horizontal positional corrections from the partial gravity gradients is approximately 300-500m along most of the route from Liverpool to Toulouse, whereas the unaided INS continues to drift, as shown in Figure~\ref{radial_error_example}. In the example shown in the figure, the interferometers are limited by the shot noise in the interference measurements. However, the current generation of gravity gradient cold atom sensors also has appreciable phase noise. So, to address this concern, we now examine the performance of the system as the level of phase noise is increased. We calculate the radial error profile for the route for 50 simulated flights with different levels of phase noise, and then average over the time of the flight. The results showing the mean radial errors for different levels of phase noise are shown in Figure~\ref{radial_error_vs_phasenoise}. We see that the horizontal accuracy of the position fixing method does degrade as the standard deviation of the phase noise approaches 10 mrad ($1\sigma$), but the errors are still significantly lower than the errors accumulated by the unaided INS navigation solution. Even for phase noise with a standard deviation of 15mrad, the mean position error is still less than 1.5km, although the standard deviation of the position fixing method does increase significantly as the phase noise increases.

\subsection{Sensitivity to Failed Measurements}\label{sec5c}

In addition to shot noise and phase noise, cold atom interferometers can also be sensitive to vibrations and other issues associated with moving sensitive equipment from the laboratory out into the field. The use of two interferometers in a gravity gradient configuration, with a common Raman laser reference, removes the worst sensitivities of the sensor to vibrations in the vertical direction, but it is possible for changes in the horizontal motion of the sensor (either sudden jerks or large vibrations) to cause the cold atoms to move outside the Raman beam and the interference measurement fail as a result. In the interferometer model used in this paper, we aim to model this by adding a random walk to the horizontal motion of the atoms during the measurement period, and we provided a failed measurement and no partial gravity gradient measurement whenever the atoms move out of the beam during the measurement cycle. We can also add in a probability for a failed measurement as a parameter. When a failed measurement occurs, the particle weights are not updated, equation (\ref{reweight}). 

\begin{figure}[t] 
   \centering
   \includegraphics[width=1.0\hsize]{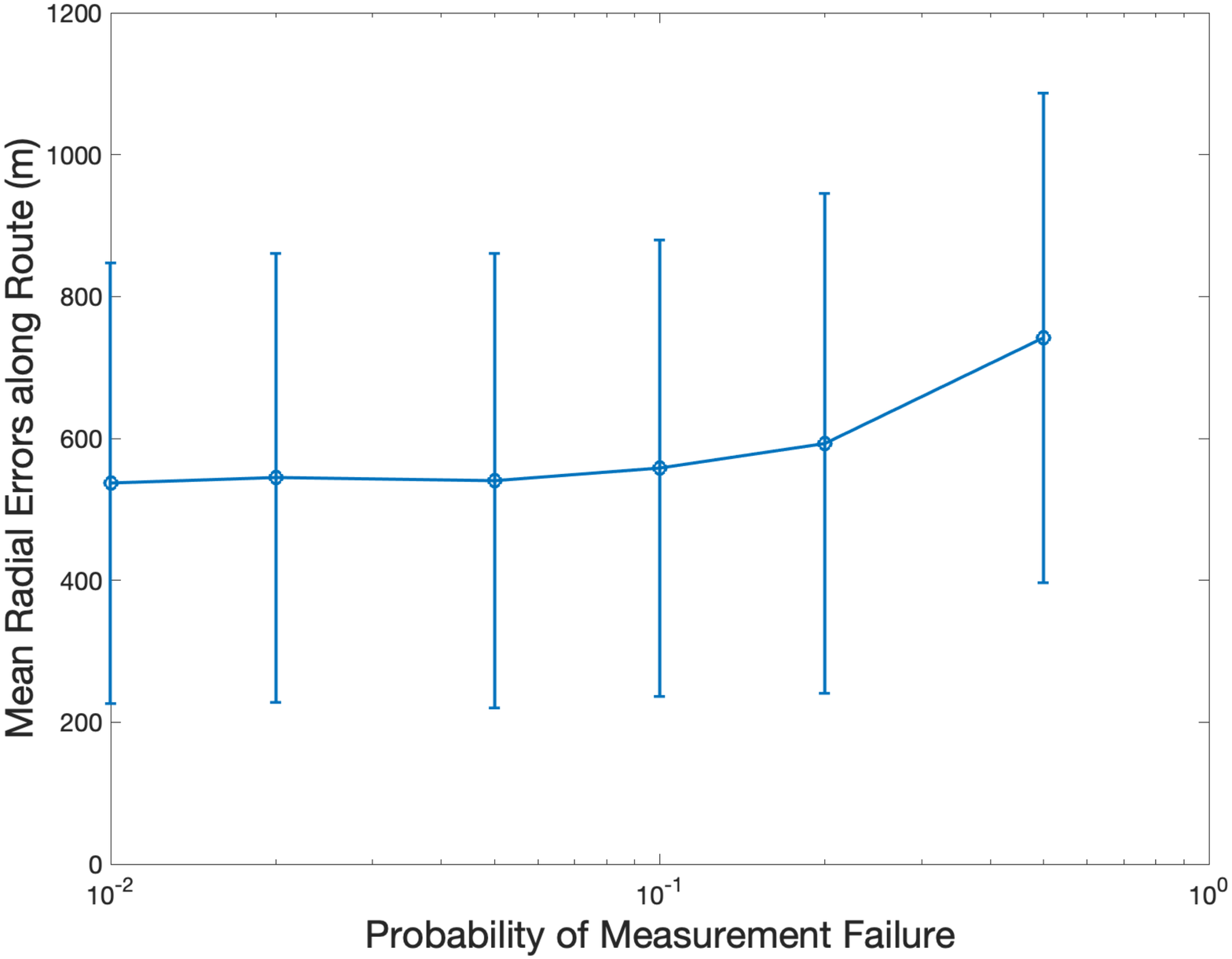} 
   \caption{Mean radial position error over 50 simulated flights and averaged along route as a function of the probability of a failed measurement in the interferometers, with error bars ($1\sigma$). (Note the log scale on the x-axis).}
   \label{radial_error_vs_failure_rate}
\end{figure}

Figure~\ref{radial_error_vs_failure_rate} shows the accuracy of the position fixing as a function of the probability of a failed measurement with an interferometer phase noise of 5 mrad ($1\sigma)$. We see that the performance of the proposed method is very good when the probability of failures is less than 20\% and only shows significant degradation when the probability of a failed measurement is increased to more than 20\%.

The fact that the proposed position fixing method is robust to significant levels of failed measurements should mean that, although we have selected a relatively benign flight path for this paper, the method should function in situations where there are short intermittent manoeuvres by the aircraft, where measurements are lost. We also note that when there are significant errors in the navigation solution corrections, the algorithm can recover and correct itself when more suitable gravity data is available, as in the situation where two databases are used, providing gravity data with different resolutions (see section~\ref{sec5a}).

\section{Conclusions}
\label{sec:sec6}

This paper has proposed a method for position fixing using partial gravity gradient measurements generated by cold atom interferometers. We have simulated flights from Liverpool to Toulouse and generated navigation solutions for both an unaided (aviation grade) inertial navigation system and an aviation grade inertial navigation system that has position corrections generated by the proposed method -- for simplicity, the simulated flights are straight and levels flights at an altitude of 3000m, flying over the UK, the English Channel/la Manche, and central France. Two open source gravity databases were used for these simulations, SRTM2gravity (over the land) and EGM2008 (over sea areas). These were used to generate gravity gradients for both the simulated interferometers and position correction method. The unaided navigation solution drifts over time, as expected, and it accumulates a mean error of around 4km during the simulated flights. The position corrected navigation solution contains errors but these are limited to a mean value of around 300-400m for most of the cases considered in this paper, only degrading significantly when the standard deviation of the phase noise in the interferometers is in excess of 10 mrad ($1\sigma$). 

The proposed method was found to be robust to small amounts vibrational noise, causing the potential for failed measurements, and for the two different resolutions of the two different gravity databases used. The positional error was found to degrade over the English Channel/la Manche using the lower resolution EGM2008 (1 nautical mile resolution), but the errors reduce when the simulated aircraft flies back over the land and is able to use the higher resolution SRTM2gravity database (90 m resolution). 

The navigation fusion/correction method used in this paper is simple, but the aim of the paper was to demonstrate the robustness of the position correction method rather than to focus on the development of an optimal navigation processing algorithm, which will be the subject of future work. In addition, it is likely that improvements to the performance outlined in this paper would be available if the processing were to be optimised to a specific gravity gradient sensor and a wider set of flight profiles. 

\acknowledgments 
The authors would like to thank the European Space Agency (ESA) for their generous funding of this work through the Quantum Wayfinder project (Grant No: NAVISP-EL1-013).

\bibliography{report}

\end{document}